# Chiropiezoelectric Energy Harvesting from Lattice-Handedness-Controlled Selenium Nanowires

Oleksii Ohiienko[1], Wookjin Jung[1], Jihyeon Yeom[1,2*]

*1. Applied Science Research Institute, Korea Advanced Institute of Science and Technology (KAIST), Daejeon 34141, Republic of Korea*

*2. Department of Materials Science and Engineering, Korea Advanced Institute of Science and Technology (KAIST), Daejeon 34141, Republic of Korea*

*Corresponding: jhyeom@kaist.ac.kr

**Abstract**

The growth of wearable electronics, soft robotics, and Internet-of-Things systems has intensified the demand for inorganic piezoelectric materials that harvest weak biomechanical energy. Advances through composition optimization, defect engineering, strain engineering, and orientation control have improved existing electromechanical responses, but have not introduced a new mechanism for enhancing piezoelectricity. Here we propose atomic chirality engineering as a design principle for inorganic piezoelectric nanomaterials. Unlike conventional strategies that modify composition or morphology, atomic chirality alters the handedness of the crystal lattice itself, adding a degree of freedom for controlling dipole alignment, electromechanical coupling, and potentially spin-dependent transport. Using atomically chiral trigonal selenium nanowires as a model system, we show that crystal handedness alone changes piezoelectric performance at identical chemical composition. Piezoresponse force microscopy resolved a consistent enantiomeric difference, with right-handed *D*-Se nanowires reaching a higher effective piezoelectric coefficient than their left-handed counterparts, and flexible nanogenerators and self-powered acoustic sensors built from the two enantiomers produced distinct outputs under identical deformation. This work establishes atomic chirality as a distinct route for designing piezoelectric materials, one that may extend across non-centrosymmetric inorganic semiconductors for sustainable energy harvesting and wearable sensing.

## Introduction

The continuous expansion of wearable healthcare, human–machine interfaces, implantable electronics, and autonomous Internet-of-Things (IoT) devices has created an urgent demand for miniature power sources capable of harvesting ubiquitous mechanical energy. Piezoelectric nanogenerators (PENGs) directly convert mechanical deformation into electrical energy without an external power supply, which has made them one of the most promising of these technologies. Extensive research has therefore focused on discovering and optimizing inorganic piezoelectric materials, including ZnO, GaN, $BaTiO_3$, lead zirconate titanate (PZT), and $ZnSnO_3$.[1-6]

Despite decades of progress, improving the efficiency of inorganic piezoelectric materials remains increasingly challenging. Existing strategies rely predominantly on chemical composition optimization, dopant incorporation, defect engineering, crystallographic orientation control, interface engineering, nanostructuring, and strain engineering. Although these approaches have substantially improved electromechanical conversion, they rarely modify the fundamental symmetry of the atomic lattice that ultimately governs spontaneous polarization and piezoelectric coupling. This limitation motivates an important question: can an entirely different atomic degree of freedom be exploited to improve piezoelectricity?

Atomic chirality represents such an opportunity. In chiral crystals, the atomic arrangement lacks mirror symmetry and exists as two enantiomeric lattices with identical chemical composition but opposite handedness.[7-10] Recent advances in chiral inorganic materials have revealed remarkable phenomena including circular dichroism,[11] spin-selective charge transport,[12,13] Rashba-related effects,[14] asymmetric catalysis,[15] and chiral photonics.[16] These studies have focused predominantly on optical or spintronic functionality, whereas the direct role of atomic chirality in electromechanical energy conversion remains largely unexplored.

From a fundamental perspective, piezoelectricity originates from symmetry breaking.[10,17] Atomic chirality introduces a symmetry descriptor beyond composition and crystallographic orientation. Altering the handedness of an atomic helix can modify local electronic polarization, dipole alignment, charge redistribution, and spin–orbit interactions without changing the chemical identity of the material.[18-20] Atomic chirality may therefore provide an unprecedented strategy for engineering piezoelectricity at its structural origin rather than optimizing secondary structural parameters.

Elemental trigonal selenium is an ideal model system because its crystal is built from intrinsically helical covalent chains that exist as two enantiomorphic space groups.[21-23] These atomically chiral lattices allow the influence of handedness on electromechanical behavior to be examined directly, without the complications that accompany changes in composition.

Here we selectively synthesize atomically chiral selenium nanowires through chirality transfer from a molecular inducer to the inorganic lattice, and show that atomic handedness governs measurable differences in piezoelectric output.[14,24,25] Atomic-resolution microscopy confirmed that the two nanowire populations are opposite enantiomers, and circular dichroism verified their mirror-image chiroptical response. Piezoresponse force microscopy resolved a reproducible enantiomeric difference in electromechanical response, with right-handed *D*-Se nanowires reaching a higher effective piezoelectric coefficient than their left-handed counterparts, and atomistic simulations traced this difference to more collinear dipole alignment in the right-handed lattice. Flexible nanogenerators and self-powered acoustic sensors fabricated from the two enantiomers showed distinct outputs under identical mechanical loading, with the right-handed enantiomer consistently stronger. These findings support atomic chirality engineering as a framework for designing high-performance inorganic piezoelectric materials. Rather than treating chirality solely as a source of optical activity, our work shows that crystal handedness itself is a tunable structural parameter for electromechanical coupling, one that may extend to other non-centrosymmetric semiconductors for energy harvesting, wearable sensing, and multifunctional quantum materials.

## Results and discussion

To obtain selenium nanowires of defined, opposite handedness, we transferred chirality from a molecular inducer to the growing inorganic lattice. Selenium was reduced in the presence of sodium dodecyl sulfate (SDS) as a capping agent, with *L*- or *D*-cysteine added as the chiral inducer. Under the mildly acidic conditions used here, the slow reduction of selenium keeps monomer supersaturation low, a regime that favors dislocation-mediated growth of well-defined crystalline wires.[26-28] Against this background, the enantiospecific interaction between the cysteine chiral center and the selenium atoms during nucleation and growth imprints the corresponding lattice handedness.[29] *L*-cysteine yielded $P3_221$ nanowires and *D*-cysteine yielded $P3_121$ nanowires, the two enantiomorphs of trigonal selenium (**Figure 1a**).

In these crystals, selenium atoms are covalently bonded into helical chains along the c-axis with threefold screw symmetry, and neighboring chains pack hexagonally through van der Waals contacts, so that the screw sense of the chains sets the handedness of the lattice.[30] Viewed along the [120] zone axis, the HAADF-STEM images matched the simulated left- and right-handed trigonal configurations for *L*-Se NW and *D*-Se NW, respectively (**Figure 1b, c**), assigning the lattice handedness of each population directly from atomic-resolution imaging. The insets show the corresponding simulated structures (BIOVIA Materials Studio).

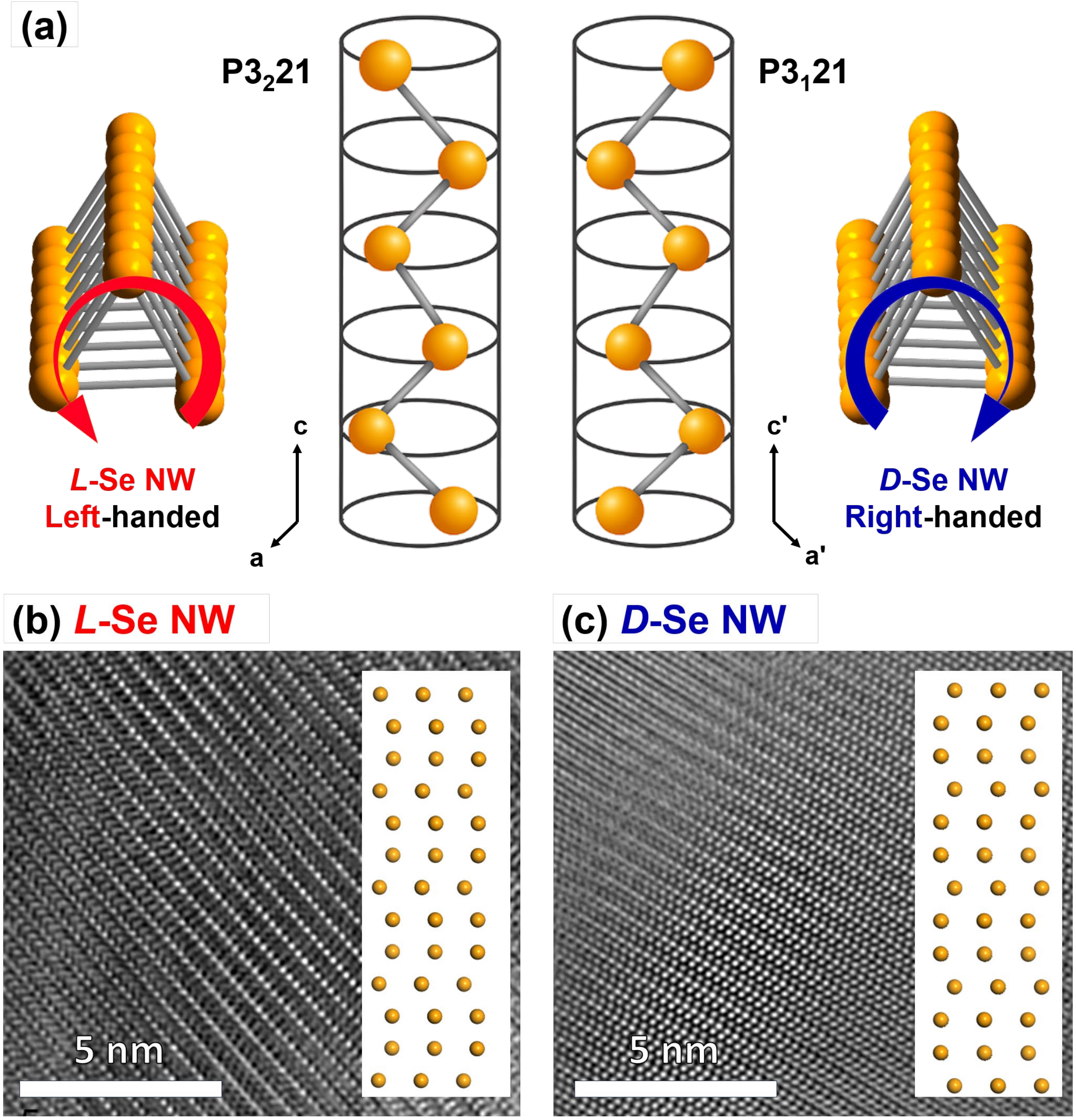

**Figure 1.** (a) Crystal structures of trigonal selenium with opposite helical handedness corresponding to the $P3_221$ and $P3_121$ space groups. (b, c) HAADF-STEM image of *L*-Se NW and *D*-Se NW respectively. Insets: simulated atomic arrangements of $P3_221$ and $P3_121$ Se crystals.

Having resolved the atomic-scale handedness by HAADF-STEM, we turned to the overall size and morphology of the wires using scanning electron microscopy (SEM) and transmission electron microscopy (TEM). Across 100 randomly selected wires, the diameters ranged from 30 to 80 nm and the lengths from 600 to 1200 nm (**Figure 2a, b)**. TEM further showed straight, smooth-surfaced wires with a uniform diameter of ~50 nm, consistent with the SEM measurements (**Figure 2c, d**). Beyond morphology, X-ray diffraction confirmed that the wires were phase-pure trigonal selenium, with all peaks indexed to the trigonal lattice (a = 4.36 Å, c = 4.97 Å, ICDD 06-0362) and no $SeO_2$ or other impurity reflections (**Supplementary Figure S1**).

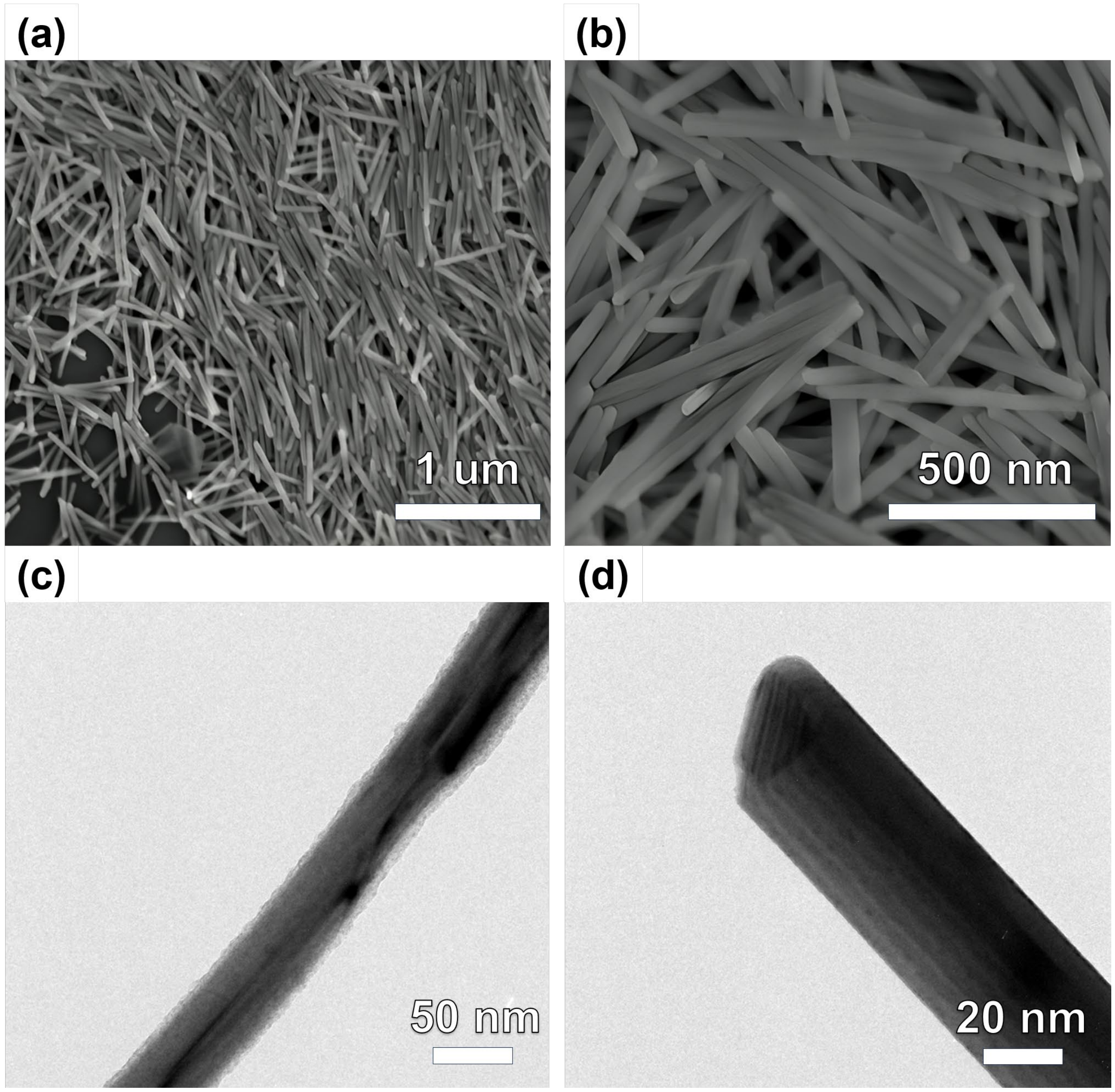

**Figure 2.** Structural and morphological characterizations of the chiral Se NWs that were prepared by the standard procedure. (a-b) SEM images of *L*-Se NWs, (c-d) TEM images of *L*-Se NWs.

To complement the atomic-scale handedness resolved by HAADF-STEM with an ensemble-level optical measurement, we measured circular dichroism (CD). CD resolved mirror-image spectra for the two enantiomers across the UV-to-NIR region (250–800 nm). *L*-Se NWs (red) showed a strong positive peak near 450 nm and a negative peak near 750 nm, and *D*-Se NWs (blue) showed the mirror-image spectrum (**Figure 3a**). Since the CD of cysteine itself is confined to the UV, the mirror symmetry of these broadband spectra confirms that the chirality originates from the selenium lattice rather than from residual molecules, and that molecular chirality was transferred to the nanowires. The dissymmetry factor (*g*-factor) reached 0.08 (**Figure 3b**), consistent with a substantial enantiomeric excess in the as-grown wires.

Having established the opposite handedness of the nanowires, we next sought to determine whether this asymmetry produces a measurable electromechanical difference. The helical, non-centrosymmetric lattice offers a direct test of how atomic handedness governs piezoelectricity. To map local polarization, we performed PFM on both *L*- and *D*-Se NWs with a conductive gold-coated cantilever.[31,32] A DC bias swept from −10 to +10 V produced well-defined butterfly-shaped amplitude–voltage loops in all samples (**Figure 3c-f**). Each loop was accompanied by a phase reversal of approximately 250°, indicating reversible polarization switching. Together, these amplitude and phase responses are consistent with piezoelectric behavior, though a minor electrostatic contribution cannot be excluded.

Quantitatively, the piezoelectric coefficient $d_{33}$ was calculated using Equation (1):

$$d_{33} = \frac{\text{Amplitude (V)} \times \text{deflection sensitivity (nm V}^{-1}\text{)}}{\text{vertical deflection gain } (\approx 16) \times \text{Applied AC bias (V)}} \quad (1)$$

The measured $d_{33}$ was 5.9 pm $V^{-1}$ for *D*-Se and 5.2 pm $V^{-1}$ for *L*-Se NW, a clear and distinct difference favoring the right-handed enantiomer. This handedness-dependent response is consistent with more effective dipole alignment in the *D*-Se NW lattice. Consistent with this, the brighter high-response domains in PFM amplitude maps were more uniformly distributed across

*D*-Se NW films, whereas *L*-Se showed weaker, spatially scattered signals (**Supplementary Figure S2**). This pattern reflects more coherent dipole alignment in the *D*-Se NW lattice, where the helical chain axis and the polarization axis are collinear and neighboring dipoles interfere less destructively. Collectively, these measurements show that atomic handedness translates into a clear, handedness-dependent electromechanical response at the nanoscale.

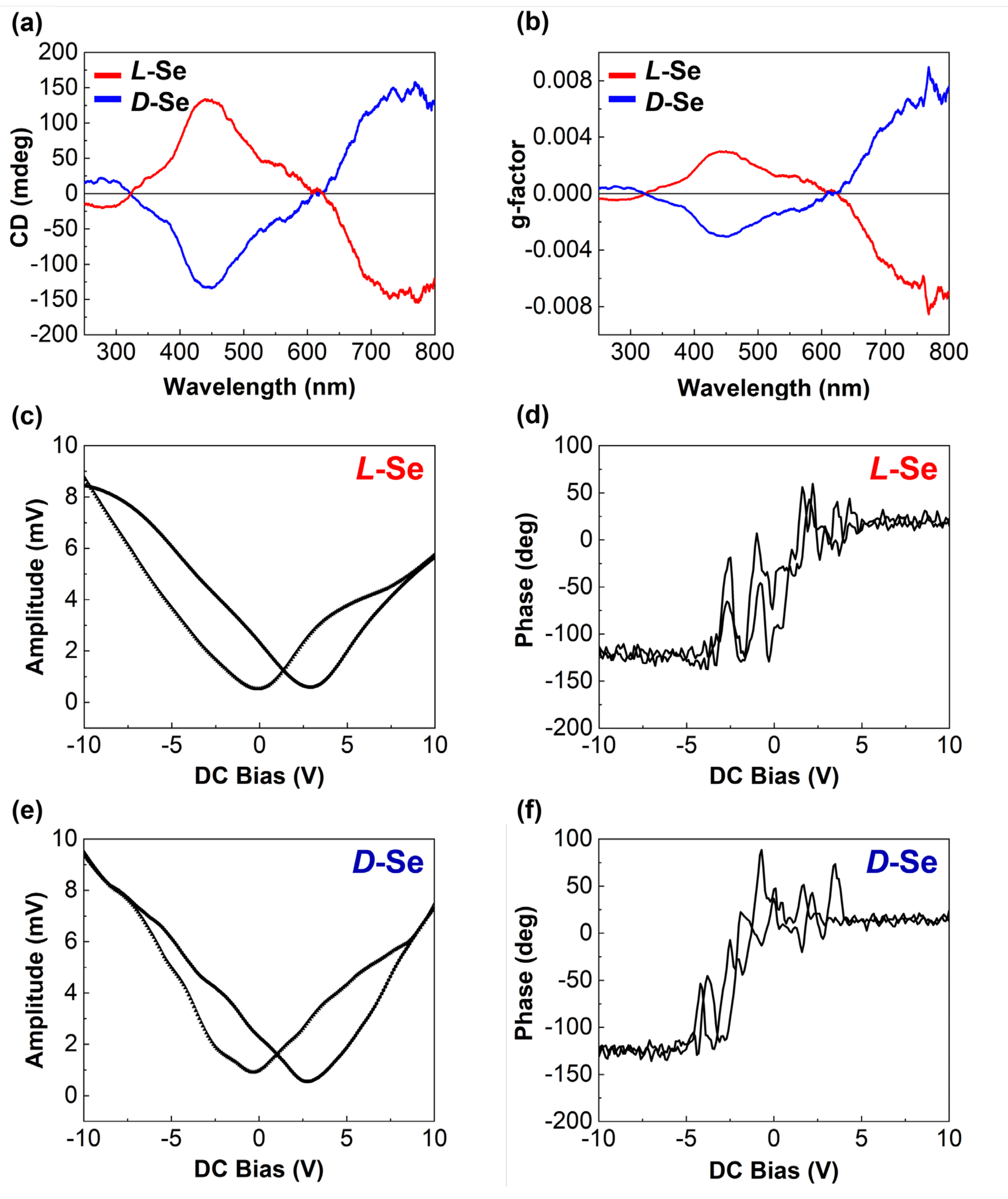

**Figure 3.** (a) Circular dichroism (CD) spectra and (b) corresponding g-factor spectra of the *L*-Se NWs and *D*-Se NWs. Amplitude-voltage and phase-voltage curves of (c,d) *L*-Se NWs and (e,f) *D*-Se NWs.

To trace the microscopic origin of this response, we performed atomistic simulations of both enantiomers under bending and compression, two representative modes of mechanical deformation. These simulations track how atomic-scale handedness shapes dipole alignment and polarization evolution as the lattice deforms.

In the undeformed state, both enantiomers showed the characteristic helical chains along the c-axis, with each selenium atom covalently bonded in a threefold screw sequence. Applying a bending stress of 1 MPa placed the outer chains under tensile strain and the inner chains under compression (**Figure 4a–d**). This asymmetric strain distribution broke the charge balance along the helical axis and drove a net dipole reorientation. Under bending, the polarization enhancement was markedly larger for *D*-Se NWs. In the right-handed lattice, the screw sense held the strain-induced dipoles collinear with the polar axis, so neighboring chains reinforced one another. In *L*-Se NWs, the opposite sense misaligned adjacent dipoles, partially cancelling them and reducing the net gain. The same asymmetry held under uniaxial compression (**Figure 4e–h**). Here, shorter interatomic distances increased orbital overlap, and the right-handed lattice again coupled this response more coherently to the polar axis. Imposing a strain gradient and a uniform strain, respectively, bending and compression correspond to flexoelectric- and piezoelectric-dominated regimes.

To quantify how strongly handedness affects the strain-induced polarization, we tracked the total dipole moment of each enantiomer under load (**Figure 4i**). Under 1 MPa, the supercell of *D*-Se NWs rose from ≈12 D to ≈22 D on bending, while the *L*-counterpart rose from ≈9 D to ≈16 D. Under compression, the *D*-supercell reached ≈24 D against ≈17 D for the *L*-supercell. In both modes the strain-induced polarization was substantially larger for the right-handed lattice, consistent with the more collinear alignment described above. Together with the PFM results, these simulations show that atomic handedness sets the coherence of dipole alignment under strain, accounting for the higher $d_{33}$ measured for *D*-Se NWs.

To interpret this handedness dependence, the geometry of the helical lattice should be considered. The helical chains of trigonal selenium define a screw axis. Under strain, the atomic displacements, orbital overlap, and polarization buildup evolve along this axis. The helical orbital motion of electrons can be treated as a nanoscale current path,[13,33] in analogy to the axial field of a solenoid.[16,34] Within this picture, a right-hand-rule argument places the screw sense and the polar axis in alignment for *D*-Se NWs, so that neighboring dipoles add coherently. In *L*-Se NWs, the opposite twist introduces an angular mismatch that partially cancels them. The net polarization therefore reflects how collinearly the strain-induced dipoles align along the screw axis. This trend is consistent with both the simulated dipole moments and the measured enantiomeric difference. The same screw symmetry that aligns the dipoles also shapes the electronic structure, where it manifests as a Berry-curvature contribution.[35]

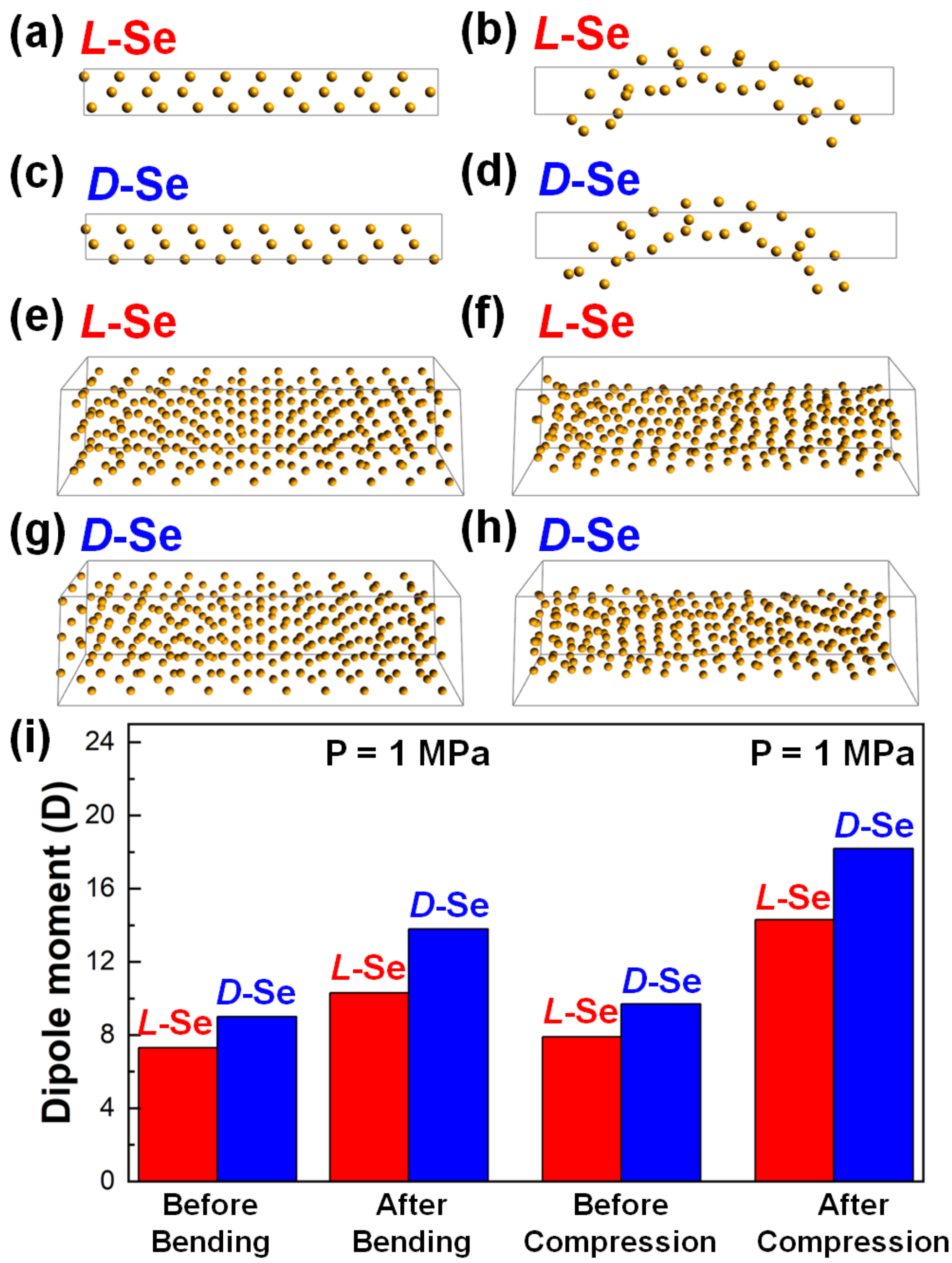

**Figure 4.** Simulation of the *L*- and *D*-Se trigonal structure with further determination of the dipole moment. (a-d) Simulative determination of the dipole moment during bending. (e-h) Determination of the dipole moment during compression. (i) Calculated dipole moments of *L*-Se and *D*-Se nanowires before and after bending or compression, highlighting the larger strain-induced dipole enhancement in *D*-Se

Our nanoscale characterization and simulations established the handedness-dependent difference at the material level, but a practical piezoelectric device operates under repeated dynamic loading rather than a quasi-static probe. We therefore asked whether the enantiomeric advantage persists at the device level, and characterized *L*- and *D*-Se PENGs under controlled cyclic loading. Each PENG combined a polyethylene terephthalate (PET) substrate, an interdigitated copper-nanowire (Cu NW) bottom electrode, a chiral Se-NW active layer, and a PET encapsulation layer. We drove the devices at 1 Hz, in bending to a curvature of ~0.316 $mm^{-1}$ and in compression under a 1 N load. Both enantiomers produced stable voltage output across the full test, confirming a durable piezoelectric response under repeated deformation. Under bending, the *D*-Se PENG reached ±50–60 mV, against ±40–45 mV for the *L*-counterpart (**Figure 5a, b**). Compression preserved this ordering, with *D*-Se at ±20–22 mV and *L*-Se at ±16–18 mV, though the enantiomeric gap was narrower than under bending (**Figure 5c, d**). Taken together, these tests confirm that the enantiomeric advantage persists at the device level, with *D*-Se the stronger performer under both bending and compression.

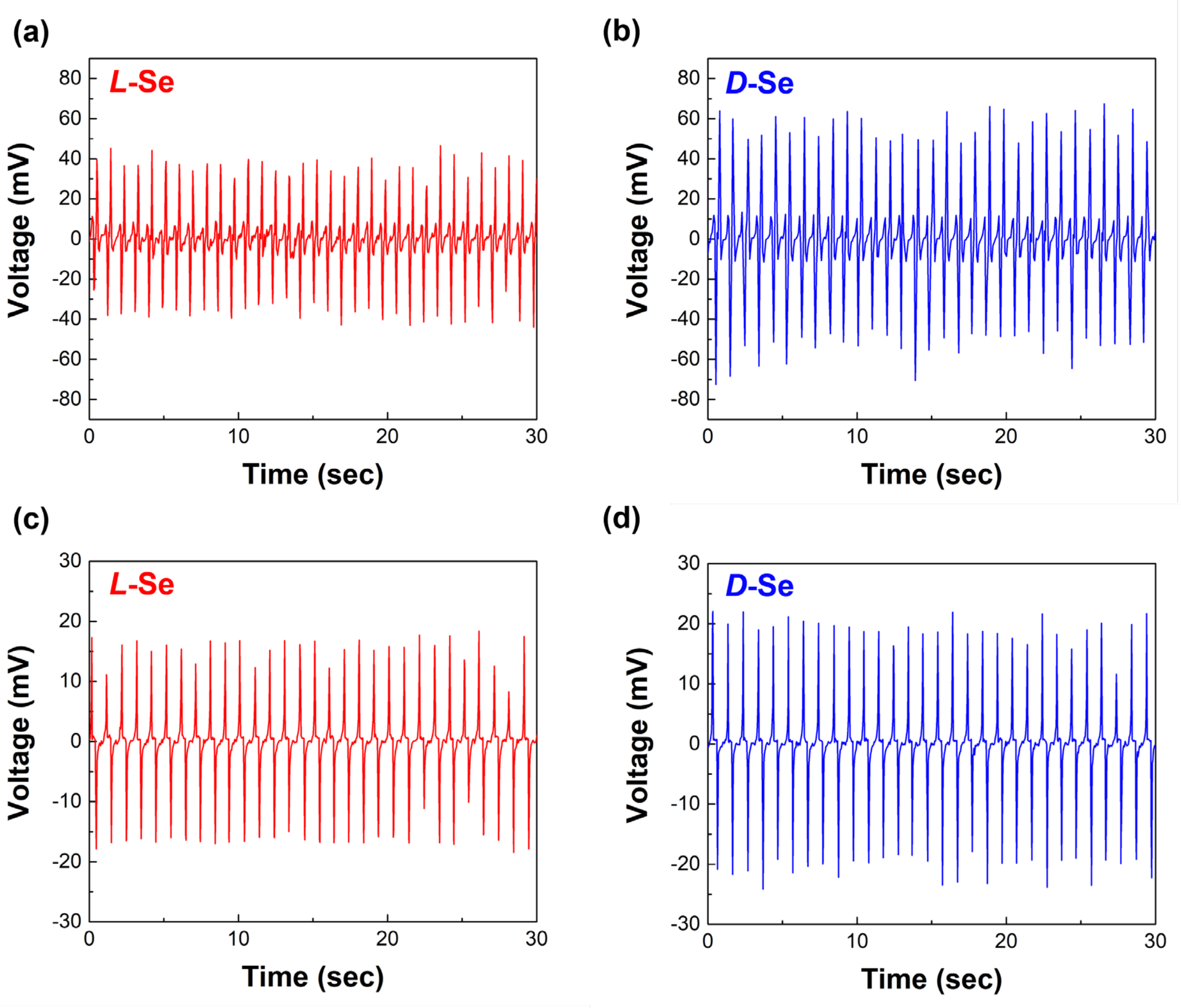

**Figure 5.** Output voltage of chiral Se NWs piezoelectric generators under mechanical deformation. (a,b) Voltage output of *L*-Se and *D*-Se devices during cyclic bending tests. (c,d) Voltage output of *L*-Se and *D*-Se devices during compression tests.

To then probe the devices under physiologically relevant deformation, we built an air-pump-driven simulator in which a soft polymer balloon inflated and deflated at 55–60 beats per minute, the typical human resting heart rate (**Figure 6a**). The rhythmic expansion and contraction generated a time-dependent strain analogous to cardiac motion. This allowed us to evaluate the PENG output in situ under physiologically realistic loading. Each simulated heartbeat produced synchronized voltage and current pulses resembling the temporal sequence of the PQRST cardiac cycle (**Figure 6b, c**). The temporal correspondence between the mechanical actuation phases (P–Q–R–S–T) and the output waveforms further confirms the sensitivity of the device to dynamic mechanical deformation and its capacity to mimic bioelectrical signals. While both enantiomeric devices gave stable, repeatable signals, the difference lay in their magnitude. The *D*-Se PENG reached 9–10 mV and 12–14 μA, whereas the *L*-Se device delivered only 6–7 mV and 9–10 μA. The right-handed enantiomer therefore produced about 30–40% higher voltage and current.

Having established the single-device response, we next asked how far the output could be scaled. Stacking multiple Se NW active layers offers a direct route, since coherently aligned dipoles should add up across layers. Across single-, double- and triple-layer devices (**Figure 6d, e**), both voltage and current rose nearly linearly with the number of Se NW layers, indicating additive polarization. In every configuration the *D*-Se PENG led, reaching ~12.5 mV and ~18 μA in the triple-layer device with stable, low-noise signals over hundreds of cycles. This advantage indicates that the right-handed lattice of *D*-Se aligns dipoles more collinearly with the strain axis, enabling higher voltage and current output. At the device scale, the *D*-Se advantage mirrors the difference measured by PFM.

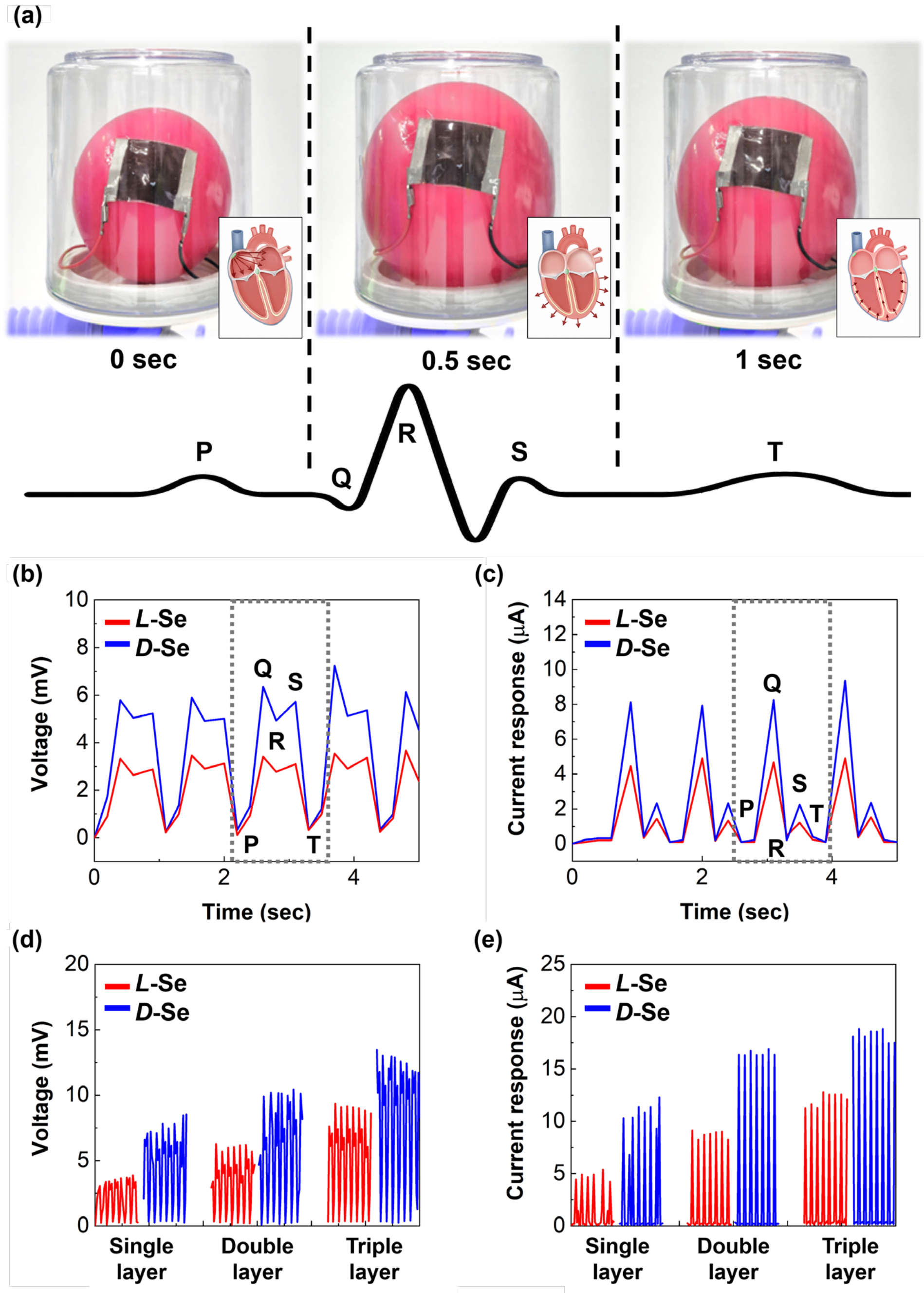

**Figure 6.** Chirality-dependent piezoelectric output of Se nanowire nanogenerators under simulated heartbeat conditions. (a) Schematic and photographs of flexible piezoelectric nanogenerators (PENGs) based on chiral Se NWs integrated onto a polymer balloon for heartbeat simulation, with insets showing the corresponding cardiac phases. (b) Time-dependent voltage and (c) current outputs of *L*-Se NW (red) and *D*-Se NW (blue) PENGs under cyclic deformation, where the dashed box marks one full P–Q–R–S–T cardiac cycle. (d) Time-dependent voltage and (e) current outputs of single-, double-, and triple-layer PENGs, showing layer-dependent enhancement and the consistent superiority of *D*-Se NWs over *L*-Se NWs in both voltage and current.

We then applied the same chiral nanowires to acoustic sensing. Specifically, we fabricated a flexible multiband sensor by spray-coating chiral Se NW strips of graded thickness (1–6 μm) onto a transparent polymer substrate and contacting them with separate silver electrodes (**Figure 7a**). Placed beside an artificial ear model, the sensor was driven by periodic sound from a calibrated speaker (**Figure 7b**). The acoustic pressure bent and compressed each Se NW layer, converting the sound into a piezoelectric voltage.

For an acoustic sensor, a narrow frequency response is a practical limitation, since real sound spans a wide spectrum. A meaningful evaluation therefore requires testing across a broad frequency range. We recorded the time-dependent output voltages of the *L*- and *D*-Se NW-based sensors across four representative frequency bands, 20 Hz–4 kHz, 4–8 kHz, 8–12 kHz, and 12–16 kHz (**Figure 7c–f**). In every band, the sensors with the *D*-Se NWs consistently produced higher voltage amplitudes than the sensors with *L*-Se NWs under identical acoustic excitation. In the low-frequency range (20 Hz–4 kHz), the *D*-Se NWs generated peaks of ~10 mV compared with ~7 mV for *L*-Se NWs. This relative advantage persisted across the higher-frequency bands. The consistent asymmetry indicates that the more collinear dipole alignment of the right-handed lattice sustains a larger piezoelectric response, even under fast dynamic strain.

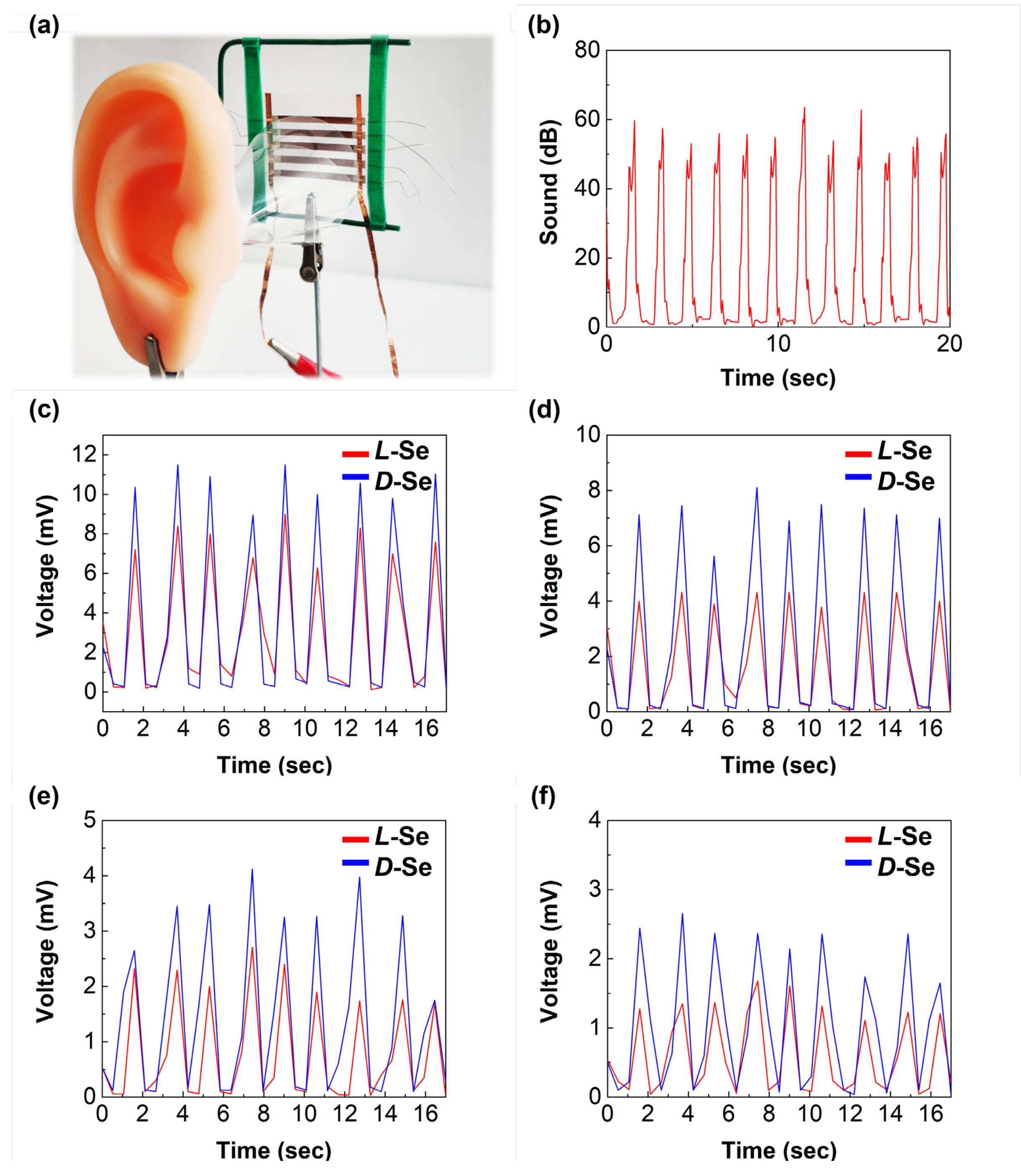

**Figure 7.** Chirality-dependent multiband acoustic sensing using Se NW-based piezoelectric devices. (a) Photograph of the artificial ear setup used for acoustic measurements with chiral Se NW sensors. (b) Input sound waveform (60 dB siren tone) recorded over time. (c–f) Time-resolved voltage outputs of *L*-Se (red) and *D*-Se (blue) acoustic sensors at frequency bands of 20 Hz–4 kHz, 4–8 kHz, 8–12 kHz, and 12–16 kHz, respectively.

## Conclusion

We have shown that the handedness of an atomic lattice, on its own, dictates the piezoelectric response of a material at fixed chemical composition. In atomically chiral trigonal selenium nanowires, circular dichroism and atomic-resolution microscopy confirmed two enantiomers of opposite handedness. Piezoresponse force microscopy then resolved a consistent difference in their electromechanical response, which atomistic simulations traced to more collinear dipole alignment in the right-handed lattice. Although the mechanical stimulus itself carries no handedness, the two enantiomers converted it into distinct piezoelectric outputs, isolating handedness as the active variable. Flexible nanogenerators and self-powered acoustic sensors built from the same nanowires carried this atomic-scale asymmetry through to device output, with the right-handed enantiomer consistently stronger. These results establish atomic chirality engineering as a design principle for inorganic piezoelectric materials, one that acts at the structural origin of piezoelectricity rather than optimizing composition or morphology. Because handedness reflects the symmetry of the lattice alone, the same principle should extend to other non-centrosymmetric inorganic semiconductors, opening a new route toward chirality-engineered energy harvesting, wearable sensing, and multifunctional quantum materials.

## Materials and Methods

### Materials

Selenium dioxide ($SeO_2$, ≥99.9%), *L*-cysteine (*L*-Cys, 98%), *D*-cysteine (*D*-Cys, 99%), sodium hydroxide (NaOH, 95%), ethanol (99%), Isopropanol (99%), hydrochloric acid (37%) were purchased from Sigma-Aldrich and used without further purification. Sylgard 184 were purchased from Dow. All water used was ultrapure from a Millipore system (18.2 MΩ·cm).

### Preparation of atomically chiral *L/D*-Se NWs

Atomically chiral selenium nanowires were synthesized via chemical reduction method for 3 h in a 40 °C water bath. First, $SeO_2$ solution (0.1 M) was added to 4.5 mL of Milli-Q water. The reaction solution was then mixed with 1 mL of Cys solution (0.1 M) (*L*-Cys, *D*-Cys for *L*-Se, *D*-Se) and 0.5 mL of SDS solution (0.1 M). Finally, 7 mL of ethanol were added into the mixture. During the reaction, the color of reaction solution was changed from orange to brick red indicating the transition from amorphous Se nanoparticles to trigonal Se nanowires. The final products were washed and collected through centrifugation with ethanol and water. After synthesis and washing, the nanowires were briefly treated with 5% acetic acid in deionized water for 20 s at room temperature to remove residual surface organics prior to device fabrication. The chiral Se nanowires were then redispersed in ethanol/IPA for further characterization.

### Fabrication of Cu-NWs

In a typical procedure, 70 mL of deionized water was mixed with 0.35 g of $CuCl_2 \cdot 2H_2O$ (or alternatively, 0.20 g of CuCl) and stirred for 30 minutes to achieve complete dissolution.

Subsequently, 0.5 g of fructose was introduced as the primary reducing agent, and the mixture was stirred for an additional 30 minutes to ensure homogeneity. Following this, 1.6 g of HDA was added to the solution to serve as a surface-capping molecule, promoting the anisotropic growth of nanowires. Afterward, 0.05 g of potassium bromide (KBr) was added as an auxiliary agent to assist in controlled reduction and stabilization of Cu nuclei. The final reaction mixture was transferred to an oil bath and maintained at 115 °C for 24 hours to complete the reduction process and promote nanowire growth.

**Fabrication of the *L/D*-Se NWs PENG**

The Cu-NWs/isopropyl alcohol solution, with a concentration of 0.2-0.3 mg/mL, was sprayed onto Polyimide substrate using a spray gun to form a uniform conductive layer. Afterward, the *L/D*-Se-NWs layer was deposited onto the Cu-NWs network by spray coating. Finally, a second PDMS layer was spin-coated over the Se nanowires, cured, and left to set overnight. The PDMS film was prepared by spin-coating a mixture of Sylgard 184 base and curing agent (in a 10:1 weight ratio), followed by curing at 90 °C overnight. PDMS was chosen as the substrate, polymer matrix, and encapsulation layer for the device, due to its excellent stretchability and transparency.

**Fabrication of the multi-band acoustic sensor**

To fabricate a flexible acoustic sensor, a Cu-NWs solution in isopropyl alcohol, with a concentration of 0.2–0.3 mg/mL, was spin-coated onto a thin PET film (5 μm) at 2000 rpm for 60 seconds, repeated four times. Chiral Se nanowires were thoroughly cleaned with alcohol and linearly spray-coated onto the PET substrate to form five distinct bands. Following the spray-coating process, the film was rinsed once with a low-concentration acetic acid solution to remove residual dust and surface contaminants. Five pairs of Ag electrodes were then connected to the individual chiral Se-NWs bands. The assembled film was subsequently cleaned with isopropyl alcohol and encapsulated with a polyurethane stabilization layer to protect against mechanical damage.

**Characterizations**

Morphology and structural characterization of the as-synthesized chiral Se nanowires were performed using a field emission scanning electron microscope (Hitachi S-4800), a Field Emission TEM microscope (FEI Company, Tecnai G2 F30 S-Twin), and a wide-angle X-ray diffractometer with 9 kW X-ray generator (RIGAKU, SmartLab Diffractometer). Circular Dichroism spectra of chiral Se nanowires were prepared using a Circular Dichroism Spectropolarimeter (Jasco Inc, CD J-1700). The current/voltage outputs of the *L*/*D*-Se-PENG were measured using an analyzer (Keithley 2700).

Supporting Information

# Chiropiezoelectric Energy Harvesting from Lattice-Handedness-Controlled Selenium Nanowires

Oleksii Ohiienko[1], Wookjin Jung[1], Jihyeon Yeom[1,2*]

*1. Applied Science Research Institute, Korea Advanced Institute of Science and Technology (KAIST), Daejeon 34141, Republic of Korea*

*2. Department of Materials Science and Engineering, Korea Advanced Institute of Science and Technology (KAIST), Daejeon 34141, Republic of Korea*

*Corresponding: jhyeom@kaist.ac.kr

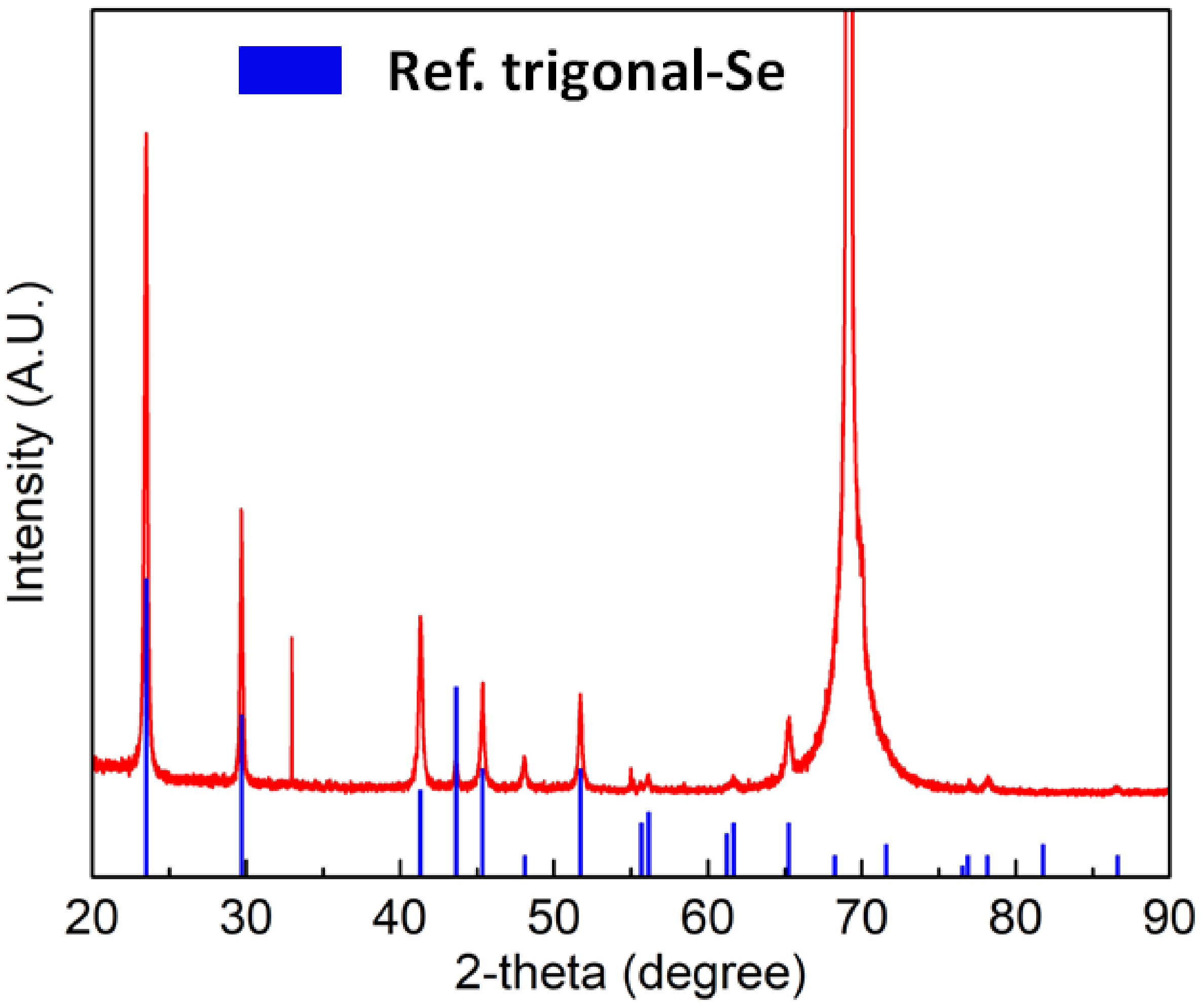

**Figure S1.** XRD patterns of the as-synthesized chiral Se nanowires.

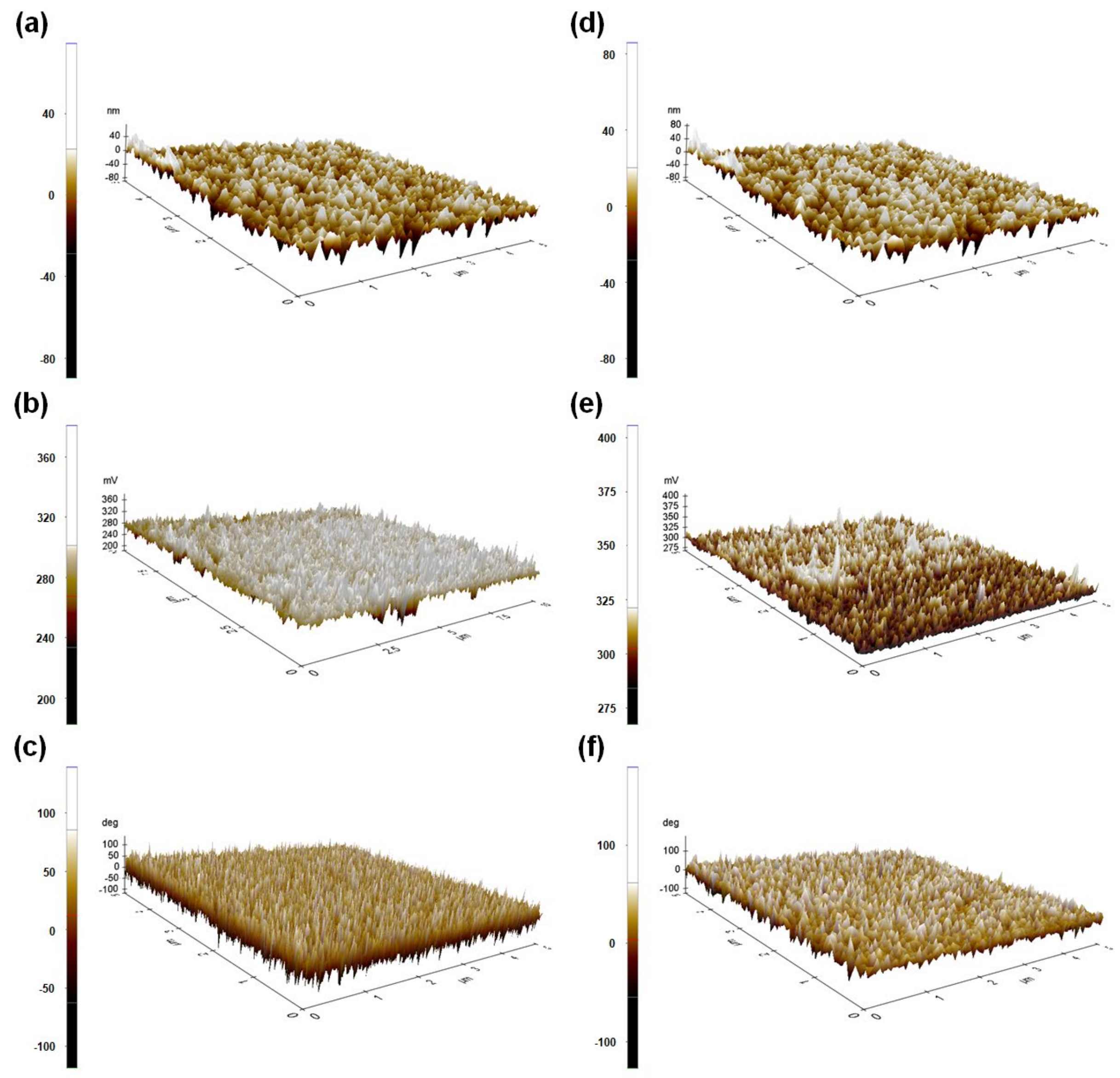

**Figure S2.** Topography, amplitude, and phase images of (a–c) *L*-Se NWs and (d–f) *D*-Se NWs measured at a driving bias of 3 V.